\documentstyle[12pt]{article}
\def\ZZZ{{\hbox{ Z\kern-1.6mm Z}}}

\newcommand{\MM}{{\cal M}}

\newcommand{\LL}{{\cal L}}

\newcommand{\wt}{\widetilde}

\newcommand{\SSS}{{\cal S}}

\newcommand{\be}{\begin{equation}}
\newcommand{\ee}{\end{equation}}
\newcommand{\ben}{\begin{eqnarray}\displaystyle}
\newcommand{\een}{\end{eqnarray}}
\newcommand{\refb}[1]{(\ref{#1})}
\newcommand{\p}{\partial}

\def\one{{\hbox{ 1\kern-.8mm l}}}
\def\zero{{\hbox{ 0\kern-1.5mm 0}}}

\begin{document}

\baselineskip 6mm
\renewcommand{\thefootnote}{\fnsymbol{footnote}}

\newcommand{\nc}{\newcommand}
\newcommand{\rnc}{\renewcommand}

%\headheight=0truein
%\headsep=0truein
%\topmargin=0truein
%\oddsidemargin=0truein
%\evensidemargin=0truein
%\textheight=9truein
%\textwidth=6.5truein

\rnc{\baselinestretch}{1.44}    % 1.5 spacing btwn text lines
\setlength{\jot}{7pt}       % spacing btwn the rows of an eqnarray
\rnc{\arraystretch}{1.64}   % spacing btwn the rows of a non-eqn array

%%%%%%%%%%%%%%%%%%%%%%%%%%%%%%%%%%%%%%%%%%%%%%%%%%%%%%%%%%%%%%%%%
%                                                               %
%                NEW COMMANDS AND MACROS                        %
%                                                               %
%%%%%%%%%%%%%%%%%%%%%%%%%%%%%%%%%%%%%%%%%%%%%%%%%%%%%%%%%%%%%%%%%

%%%%% Simplify some frequently used LaTeX commands %%%%%

%\nc{\be}{\begin{equation}}

%\nc{\ee}{\end{equation}}

\nc{\bea}{\begin{eqnarray}}

\nc{\eea}{\end{eqnarray}}

%\nc{\ben}{\begin{eqnarray*}}

%\nc{\een}{\end{eqnarray*}}

\nc{\xx}{\nonumber\\}

\nc{\ct}{\cite}

%\nc{\la}{\label}

\nc{\eq}[1]{(\ref{#1})}

\nc{\newcaption}[1]{\centerline{\parbox{6in}{\caption{#1}}}}

\nc{\fig}[3]{

\begin{figure}
\centerline{\epsfxsize=#1\epsfbox{#2.eps}}
\newcaption{#3. \label{#2}}
\end{figure}
}

%%% Double line letters %%%

\def\IR{{\hbox{{\rm I}\kern-.2em\hbox{\rm R}}}}
\def\IB{{\hbox{{\rm I}\kern-.2em\hbox{\rm B}}}}
\def\IN{{\hbox{{\rm I}\kern-.2em\hbox{\rm N}}}}
\def\IC{\,\,{\hbox{{\rm I}\kern-.59em\hbox{\bf C}}}}
\def\IZ{{\hbox{{\rm Z}\kern-.4em\hbox{\rm Z}}}}
\def\IP{{\hbox{{\rm I}\kern-.2em\hbox{\rm P}}}}
\def\IH{{\hbox{{\rm I}\kern-.4em\hbox{\rm H}}}}
\def\ID{{\hbox{{\rm I}\kern-.2em\hbox{\rm D}}}}

%%%%% Roman pont in math

\def\Tr{{\rm Tr}\,}
\def\det{{\rm det}}
\def\sv{{\cal v}}

%%%%% Special Letters

\def\vare{\varepsilon}
\def\barz{\bar{z}}
\def\barw{\bar{w}}

\begin{titlepage}
%---------------- preprint number ---------------
\hfill\parbox{4cm}

\begin{flushright}
{IPM/P-2006/003\\
IP/BBSR/2006-1}\\
hep-th/0604028\\
\end{flushright}

\vspace{25mm}
\begin{center}
%------------------------ title ------------------------
{\Large{\bf Born-Infeld Corrections to the Entropy Function\\
\vskip 0.2cm
of Heterotic Black Holes} } 

\vspace{12mm}
%---------------- authors and addresses ----------------

B. Chandrasekhar\footnote{Email: chandra@iopb.res.in}\footnote{
Present address: Institute of Physics, Bhubaneswar 751 005, India}\\

{\small {\em Institute for Studies in Theoretical Physics and
Mathematics (IPM),}} \\
{\small {\em P.O.Box 19395-5531, Tehran, Iran}}
\\

\end{center}

\thispagestyle{empty}

\vskip 2cm

%----------------------- abstract ----------------------

\centerline{\bf Abstract} \bigskip

We use the black hole entropy function to study the effect 
of Born-Infeld terms on the entropy of extremal black holes in 
heterotic string theory in four dimensions. We find
that after adding a set of higher curvature terms to the effective
action, attractor mechanism works and Born-Infeld terms 
contribute to the stretching of near horizon
geometry. In the $\alpha' \rightarrow 0$ limit, 
the solutions of attractor equations for moduli fields and the 
resulting entropy, are in conformity with the ones for standard two 
charge black holes.

\vfill \eject

\baselineskip=18pt

\end{titlepage}

\tableofcontents

\section{\bf Introduction}

String theories in their low-energy limit give rise to effective models
of gravity in higher dimensions which involve an infinite series of
higher curvature terms. The higher curvature corrections to the 
effective action considered of interest come from two sources, to wit,
$\alpha'$ corrections and string loop corrections in a
given string theory. There has been considerable work on understanding
the role of these higher curvature terms from various points of 
view, especially with regard to black hole 
physics~\cite{Boulware:1985wk}-\cite{0412287}.
One of the contexts which has received considerable attention,
is the effect of higher curvature terms
on the Black Hole entropy~\cite{9307038}-\cite{9502009}.  
Intriguingly, the inclusion of $\alpha'$ corrections in the 
low energy effective action leads to a 
resolution of naked singularities for certain classical 
solutions in string
theory~\cite{0410076} and then the solutions describe black holes with
a regular horizon. In fact, the effect of these higher derivative 
corrections for the near horizon geometry were anticipated 
from the `stretched horizon' proposal in string 
theory~\cite{9504147}-\cite{0411255}. As a consequence of 
these corrections,
the black hole entropy also comes out to be finite.

In the presence of higher curvature terms, there are deviations from
Bekenstein-Hawking area law. One of the hallmarks of string theory
is the exact matching of black hole entropy computed from
microscopic counting of the degeneracy of 
string states~\cite{9601029,9711053,9711067} and
the macroscopic calculation from the solution carrying the same 
charges, even in the presence of higher curvature 
interactions~\cite{9711053}-\cite{0412287}. Even more so, in some 
cases, the sub leading corrections to the entropy of certain black  
holes, which already have non-zero horizon area classically, have also been 
computed. These computations further the proposals
relating extremal black holes 
and elementary string states~\cite{9202014,9504147,SenIPMtalk}.

An important 
ingredient of the black hole entropy computations involving 
higher derivative terms, is the calculational tool provided by 
supersymmetry. In particular, a rich structure has emerged
out of the study of BPS black holes in ${\cal N}=2$ supersymmetric
string theories. One of the remarkable features of black holes in 
these theories is the attractor mechanism, whereby, the scalar 
fields are drawn to fixed points as they approach the 
horizon~\cite{9508072}-\cite{9602136}.
In other words, the near horizon values of the scalar fields are 
independent of the asymptotic values of these moduli fields and
they get fixed in terms of the charges carried by the black hole.
As a result, the black hole solution and the associated entropy near
the horizon is determined completely in terms of the conserved 
charges akin to the black hole. The entropy of these black holes
has been calculated both in the supergravity approximation and even
after the inclusion a certain set of higher curvature terms in 
the generalized prepotential. The
result agrees with the microscopic counting done on the brane 
system they describe.

More recently, a {\it generalized attractor mechanism} has been formulated 
without making explicit use of supersymmetry and in the presence of higher
derivative terms(see~\cite{0009234} for discussion with supersymmetry). 
There have been various approaches\footnote{For earlier work where 
attractor mechanism was understood on the basis of regularity 
of the geometry and various moduli fields near the 
horizon see~\cite{9702103}.} to this 
subject~\cite{9812082,0405146}-\cite{0601016}. 
In particular, an {\em `entropy function'} formalism 
was put forward in~\cite{0506177} to 
study the role of higher derivative terms on the entropy of 
extremal black holes from the near horizon geometry, using 
Wald's entropy formula. Also, a {\em c-extremization} procedure was  
proposed in~\cite{0506176} assuming a near horizon geometry which 
contains an $AdS_3$ factor. 

In the entropy function formalism, to a first
order approximation, it is assumed that the structure of 
near horizon geometry of the extremal black holes in this theory
remains $AdS_2\times S^2$ in four dimensions, even after 
the inclusion of higher derivative terms in the
action~\cite{0009234,0505122}. Based on this assumption, 
the black hole entropy is then
understood as the Legendre transform of Lagrangian 
density~\cite{0506177,SenToronto05} multiplied by $2\pi$,
confirming the proposal of~\cite{0405146}.
Using the above approach, the entropy of two, four and
eight charge black holes in heterotic string theory was 
calculated in~\cite{0508042}.
Intriguingly, it is possible to perform the calculation in a
regime, where all the higher derivative terms come purely from 
$\alpha'$ corrections. In this regime, the string loop corrections
can be neglected. Therefore in~\cite{0508042}, the corrections to
black hole entropy occurring from the inclusion of specific class of
higher derivative terms, namely, Gauss-Bonnet terms was computed. 
The Gauss-Bonnet terms appear at order $\alpha'$ in the heterotic
string effective action~\cite{plb156,9610237}. 
The extremization of `entropy function' with
respect to the moduli, then determines the generalized attractor equations.
Attractor equations fix the values of moduli in terms of the charges
carried by the black hole. The results can be generalized to include higher
Gauss-Bonnet densities, based on Lovelock type of actions~\cite{0511306}.
All the same, it is not yet clear why only a 
particular set of higher curvature corrections
should be added to a given system. A possible hint is the 
anomaly approach to black hole entropy~\cite{0506176}. For some of 
the recent developments, see~\cite{Jatkar:2005bh}-\cite{Exirifard:2006wa}.

It is to be noted that in the aforementioned works, only the higher curvature 
corrections ensuing from the gravity side were included and their effect
on the entropy was studied. However, if one is to consider the Maxwell
fields coupled to a gravitational action, which also includes string
generated corrections at higher orders in $\alpha'$, then, it is 
naturally important to consider string generated corrections to 
the electromagnetic field action as well. Such corrections come from
a coupling of abelian gauge fields to open bosonic or open 
superstrings. To be precise, it is known that, just like the Gauss-Bonnet
terms, there are Born-Infeld terms~\cite{Born:1934gh}, 
which appear as higher order $\alpha'$
corrections to the Maxwell action~\cite{plb163}-\cite{9908105}. 

However,
explicit corrections of Born-Infeld type terms on the entropy of
extremal black holes have not been computed thus far and their exact 
dependence is not known. Furthermore, it is important to study attractor
mechanism in more general situations where there are higher order gauge
and gravitational terms in the effective action. It is to be noted that the
effect of these terms on the entropy of black holes is in general hard to
compute analytically. It is the purpose of this
letter to consider the effect of a certain class of string generated 
corrections to the electromagnetic part, appearing in the low energy
limit of heterotic string theory and study their effect on the `entropy
function' and attractor mechanism. In particular, we shall modify the
Maxwell part of the heterotic string action by including a more general
Born-Infeld part minimally coupled to the dilaton 
and other scalar fields.  In fact, the
derivative independent part of the Born-Infeld action is exact to all
orders in $\alpha'$. Considering the analogy between the Gauss-Bonnet
and the Born-Infeld terms, which are on similar footing with regard
to string corrections on the gravity side and Maxwell side, 
respectively, it is important to include both these corrections 
simultaneously.

On more important grounds, since the inclusion of higher curvature terms
gives rise to a finite area of the horizon of small
black holes in heterotic string theory, it is interesting to 
explore whether the Born-Infeld terms have any role to play 
in this regard. Analogs of 
Reissner-Nordstrom Black hole solutions in 
Born-Infeld theories coupled to gravity and with general Lovelock type 
actions and dilaton, are being studied in 
literature~\cite{0004071}-\cite{0101083}. However, while working with a 
complete solution, studying the attractor mechanism requires a different
methodology, especially
in the presence of higher curvature terms in the action; for instance,
see~\cite{0507096,Chandrasekhar:2006kx}.
The entropy function formalism 
put forward in~\cite{0506177}, allows us to compute
the entropy of an extremal configuration from the near horizon 
geometry, without the explicit knowledge of a complete solution. 
Thus in this work, our approach would be to study the attractor mechanism
and entropy of such black holes in the 
near horizon limit, by directly adding the relevant terms to the action,
in the same vein as~\cite{0508042}. A complete study of attractor 
mechanism with the full solution, we leave for a future work. The near 
horizon analysis in fact already shows some interesting features, as we
will see. Notice that we are considering an 
extremal configuration, where it can be assumed that the solution near
the horizon does not get corrected by adding $\alpha'$ corrections
to the effective action. Of course, it is not assured that 
this will be the case, if a similar analysis is to be repeated for
non-extremal black holes~\cite{9908105}.

It is useful to recollect that the importance of Born-Infeld terms
in the study of extremal black holes has been emphasized by Gibbons
and Rasheed in~\cite{9506035}. It was noted 
that virtual black holes going around
closed loops may induce corrections to the Maxwell 
lagrangian, which, at the  
string tree level, may be of the Born-Infeld type, though there are other 
theories giving non-linear
electrodynamics in four dimensions. Extreme electrically charged black holes
in Born-Infeld theories exist and the `double degenerate' horizon 
occurs at $r_H = \sqrt{\frac{e^2}{2\pi}-\frac{1}{32\pi\,b^2}}$, 
where $e$ is the electric charge. 
Thus, there appears a natural bound on the electric charge
for the existence of consistent solutions of the Reissner-Nordstrom
type, albeit, with negative binding 
energy~\cite{Wiltshire:1988uq,9506035}(see also~\cite{9702087}). 
It is further interesting to address what these extremal configurations
correspond to in terms of elementary string states. A possible hint is that, 
the charge to mass ratio of the Born-Infeld black
holes closely matches the result for the Maxwell Reissner-Nordstrom
black holes~\cite{9506035}. In this paper, we attempt to compute the entropy
of such extremal configurations in heterotic string theory with Born-Infeld
corrections.

The rest of the paper is organized as follows. In section-2, we write
down the low-energy effective action of heterotic string theory in 
four dimensions with
Born-Infeld type corrections and determine the `entropy function' for 
a four charge black hole. In section-3, we study the attractor mechanism
in two charge black holes in this theory and show that the solutions of
attractor equations are consistent only after the addition of higher 
curvature terms on the gravity side. We also illustrate that 
our results are in conformity with the standard results for two 
charge black holes in heterotic string theory, 
in the $\alpha' \rightarrow 0$ limit. Our conclusions are 
summarized in section-4.

\section{\bf Entropy Function with Born-Infeld Corrections}

Our starting point is the low-energy effective action of heterotic string
theory on $\MM\times S^1\times \wt S^1$, where $\MM$ is some four manifold. We
follow the notations and conventions of~\cite{0508042} and review certain 
aspects
needed for our purposes. Thus, we work in the 
supergravity approximation, where the four dimensional 
fields relevant for studying the black hole solution
are related to the ten dimensional string metric $G^{(10)}_{MN}$,
anti-symmetric tensor field $B^{(10)}_{MN}$ and the dilaton $\Phi^{(10)}$ 
as follows~\cite{0508042,Cvetic}:
\ben \label{e6}
 &&  {\Phi = \Phi^{(10)} - {1\over 4} \, \ln (G^{(10)}_{99})}
 - {1\over 4} \, \ln (G^{(10)}_{88}) - {1\over 2}\ln V_\MM
 \, ,\nonumber \\
&&  {S=e^{-2\Phi}}\, , \qquad
  {R = \sqrt{G^{(10)}_{99}}}\, , \qquad
 \wt R = \sqrt{G^{(10)}_{88}}\, , \nonumber \\ 
&& {G_{\mu\nu} = G^{(10)}_{\mu\nu} - (G^{(10)}_{99})^{-1} \,
G^{(10)}_{9\mu}
\, G^{(10)}_{9\nu} - (G^{(10)}_{88})^{-1} \,
G^{(10)}_{8\mu}
\, G^{(10)}_{8\nu}\, , }\nonumber \\ 
&& A^{(1)}_\mu = {1\over 2} (G^{(10)}_{99})^{-1} \, G^{(10)}_{9\mu}\, ,
\qquad   {A^{(2)}_\mu = {1\over 2} (G^{(10)}_{88})^{-1} \, G^{(10)}_{8\mu}\, ,}
\nonumber \\ 
&&  {A^{(3)}_\mu = {1\over 2} B^{(10)}_{9\mu}\, , }
\qquad \qquad \quad {A^{(4)}_\mu = {1\over 2} B^{(10)}_{8\mu}}\, .
 \een
Here, $V_\MM$ is the volume of $\MM$ measured in the string metric.
The effective action governing the dynamics of these fields is:
 \ben \label{e4+}
&&\SSS = {1\over 32\pi} \int d^4 x \, \sqrt{-\det G} \,
S \, \bigg[ R_G 
+ S^{-2}\, G^{\mu\nu} \, \p_\mu S \p_\nu S -  R^{-2}
\, G^{\mu\nu} \, \p_\mu R \p_\nu R -  \wt R^{-2}
\, G^{\mu\nu} \, \p_\mu \wt R \p_\nu \wt R \nonumber \\
&& \, - R^2 \, 
G^{\mu\nu} \, G^{\mu'\nu'} \, F^{(1)}_{\mu\mu'} 
F^{(1)}_{\nu\nu'} - \wt R^2 \, 
G^{\mu\nu} \, G^{\mu'\nu'} \, F^{(2)}_{\mu\mu'} 
F^{(2)}_{\nu\nu'} - R^{-2} \,
G^{\mu\nu} \, G^{\mu'\nu'} \, F^{(3)}_{\mu\mu'}
F^{(3)}_{\nu\nu'} \xx 
&& - \wt R^{-2} \,
G^{\mu\nu} \, G^{\mu'\nu'} \, F^{(4)}_{\mu\mu'}
F^{(4)}_{\nu\nu'}\bigg] \qquad +\hbox{ higher derivative terms 
+ string loop corrections.}
\een
Here, $A^{(1)}_\mu$ and $A^{(3)}_\mu$ couple to the momentum and
winding numbers along the $x^9$ direction, whereas the fields 
$A^{(2)}_\mu$ and $A^{(4)}_\mu$ couple to the momentum and winding
numbers along the $x^8$ direction with the following 
identifications~\cite{0508042}:
\be \label{erel}
q_1={1\over 2} n, \quad q_3={1\over 2} w, \quad p_2=4\pi\wt N, \quad
p_4=4\pi \wt W\, .
\ee
where $q_1$ and $q_3$ are electric charges and $p_2$ and $p_4$ are 
magnetic charges. In~\cite{0505122}, it was shown that the black hole
solution in the limit where the $\alpha'$ corrections are important,
is completely universal and higher derivative terms give corrections
in a specific way.  In this work, we shall consider the Born-Infeld terms
and study the entropy function in the resulting theory. 

The Lagrangian density describing Born-Infeld theory in
arbitrary space-time dimensions is:
\be
{\mathcal L}_F =  {4\,b} \left\{ \sqrt{|\det G}| -
  \sqrt{|\det(G_{\mu\nu}+\frac{1}{\sqrt b}\,F_{\mu\nu})|} \right\},
\label{LBI1}
\ee
Restricting to four space-time dimensions and using the Monge gauge (or static
gauge) the determinant in the above action can be expanded
to obtain:
\begin{equation}
{\cal L}_{BI} = 4b S^{-1}\left\{1  -  \left[1 +
\frac{S^2}{2b}F^2  - \frac{S^4}
{16b^2}(F\star F)^2  \right]^\frac{1}{2} \right\},
\end{equation}
where we have also included the dilatonic coupling. 
Here $\star F$ is dual to the Maxwell tensor which does not contribute
in the following discussion.
% and  $\alpha$ is the dilaton coupling parameter. 
In the context of string
theory, the parameter $b$ is related to string tension
$\alpha^\prime$ as: $b^{-1}= (2\pi\alpha^\prime)^2$. Note that in the
$b\to \infty$ limit the action reduces to the Maxwell system. 
We continue to retain the Born-Infeld parameter, so as to make
a comparison with~\cite{0508042}. We set $\sqrt{\alpha'}=4$ 
and the Newton's constant $G_N =2$. With these conventions, 
$b^{-1} = {(32\pi)}^2$.

Let us then take the effective field theory action of heterotic string in the
supergravity limit to be:
 \ben \label{bi}
&&\SSS = {1\over 32\pi} \int d^4 x \, \sqrt{-\det G} \,
S \, \bigg[ R_G 
+ S^{-2}\, G^{\mu\nu} \, \p_\mu S \p_\nu S \quad -  R^{-2}
\, G^{\mu\nu} \, \p_\mu R \p_\nu R \xx
&&-  \wt R^{-2}
\, G^{\mu\nu} \, \p_\mu \wt R \p_\nu \wt R\, - R^2 \, {\cal L}_{BI}^{(1)}
- \wt R^2 \, {\cal L}_{BI}^{(2)} - R^{-2} \,{\cal L}_{BI}^{(3)}
 - \wt R^{-2} {\cal L}_{BI}^{(3)} \,\bigg] \nonumber \\
&& \qquad +\hbox{ higher derivative terms 
+ string loop corrections}
\een
and here the higher derivative terms and string loop corrections denote
the ones coming from gravity side. The Born-Infeld terms can be written 
in a useful form as:
\be
{\cal L}_{BI}^{i} = 4b S^{-2}\left\{1  -  \left[1 +
\frac{S^2}{2b}\, G^{\mu\nu} \, G^{\mu'\nu'} \, F^{(i)}_{\mu\mu'} 
F^{(i)}_{\nu\nu'} \right]^\frac{1}{2}\right\}.
\ee
We have adjusted the dilaton couplings and the coefficients in such a way
that in the $\alpha' \rightarrow 0$ limit, we recover the action given
in eqn. (\ref{e4+}). The particular coupling of dilaton
appearing in the Born-Infeld terms can also be motivated from $S$-duality
arguments from Type-I superstring theory. 
%Let us mention in the passing that
%we do not include any world volume gauge fields in this analysis.

Let us now consider an extremal charged black hole solution 
in this theory, in its
near horizon limit, with $AdS_2\times S^2$ as the near horizon geometry. 
Let us further assume that
the string generated corrections, do not spoil this property, 
and only give corrections to various parameters
characterizing the near horizon geometry, thereby, changing relations between
them. Thus, with all these conventions for the Born-Infeld corrections 
in the effective action (\ref{bi}), let us consider the near horizon 
geometry to be~\cite{0508042}:
\ben \label{eb1}
 {ds^2  = v_1\left(-r^2 dt^2+{dr^2\over 
r^2}\right)   +
v_2 (d\theta^2 + \sin^2\theta d\phi^2)\, , }\nonumber \\ 
 S =u_S, \qquad R=u_R , \qquad  {\wt R = u_{\wt R} }\, ,\nonumber \\
  {F^{(1)}_{rt} = e_1,} \quad F^{(3)}_{rt}=e_3, \quad
 {F^{(2)}_{\theta\phi}={p_2\over 4\pi}, }\quad F^{(4)}_{\theta\phi}
= {p_4\over 4\pi}    \, ,
\een
where $v_1,v_2$ are the radii of $AdS_2$ and $S^2$, $e_1, e_3$
are radial electric fields and $p_2, p_4$ are the radial magnetic fields,
$u_S$ stands for the dilaton and $u_R,{\wt u_R}$ characterize the radii of $S^1$ and
$\wt S^1$. All these objects are constants in the near horizon limit. 

Following the prescription in~\cite{0508042}, let us consider the function:
\be \label{f}
f(u_S, u_R,\wt u_R, v_1,v_2, e_1, e_3,p_2, p_4)   
{\equiv \int d\theta d\phi \, \sqrt{-\det G} \, \LL} \, .
\ee
The equations of motion of scalar fields and the metric, equivalent
to Einstein's equations, can be obtained
by extremizing the function 
$f$ with respect to $u_S,u_R,{\wt u_R},v_1$ and $v_2$. 
The gauge field equations are derived as follows. 
It is usual to define the tensor $G^{\mu\nu}$ as:
\be
G^{(1)\mu\nu} = -{1\over 2} 
{\partial {\mathcal L}^{(1)}_F\over\partial F^{(1)}_{\mu\nu}} = -
{S^2\,R^2\,F^{(1)\mu\nu} \over \sqrt{1+ \frac{S^2}{2b}\,F^{(1)\,2}} }\, ,
\ee
so that $G^{(1)\mu\nu}\approx F^{(1)\mu\nu}$ for weak fields (and with similar
results for other gauge fields). The tensor $G^{\mu\nu}$ satisfies
the electromagnetic field equations, the non-trivial components of which are:
\be
\partial_r(\sqrt{-\det G}\, G^{(i)rt}) = 0, \qquad \partial_r\,G^{(i)\theta\phi} =0.
\label{EOMBI}
\ee
where $i$ runs over all the gauge fields.
As in~\cite{0508042}, even for the present case, these gauge field
equations are satisfied for the background (\ref{eb1}). The entropy 
associated with this black hole can be calculated using the 
formula~\cite{9307038,9312023,9403028,9502009}:
\be \label{wald}
S_{BH} = 8\pi\, {\p \LL\over \p R_{rtrt}} \, g_{rr} \, g_{tt} \, A_H\, ,
\ee
where $A_H$ denotes the area of the event horizon. This formula can be 
brought in to a form, where the black hole entropy can be obtained from the 
Legendre transform of $f$ as~\cite{0506177}:
\be \label{e24}
S_{BH} = 2\pi\, \left( e_i \, {\p f\over \p e_i} - f\right) \, ,
\ee
where the function $f$ is defined in eqn. (\ref{f}). In our case, 
for the action given in eqn. (\ref{bi}), $f$ turns out to 
have the form:
\bea \label{fbi}
&&f(u_S,u_R,{\wt u_R},v_1,v_2,e_1,e_2,p_2,p_4) \, \xx
&&= {1\over 8} \, v_1 \, v_2
\, u_S  \left[ -{2\over v_1} +{2\over v_2} + 
4b \,  u_R^2 \, u_S^{-2}\left\{1  -  \left[1 -
\frac{u_S^2}{b}\, {e_1^2\over v_1^2} \right]^\frac{1}{2} \right\} \right. \xx
&&+ \left. 
4b \,  u_R^{-2} \, u_S^{-2}\left\{1  -  \left[1 -
\frac{u_S^2}{b}\, {e_3^2\over v_1^2} \right]^\frac{1}{2} \right\}+ 
4b \,  u_{\wt R}^2 \, u_S^{-2}\left\{1  -  \left[1   +
\frac{u_S^2}{b}\, {p_2^2 \over 16\pi^2 v_2^2} \right]^\frac{1}{2} \right\} 
\right. \xx
&&+ \left. 4b \,  u_{\wt R}^{-2} \, u_S^{-2}\left\{1  -  \left[1 +
\frac{u_S^2}{b}\,  {p_4^2 \over 16\pi^2 v_2^2}  \right]^\frac{1}{2} \right\}
\right].
\eea
Now, we would like to take a Legendre transform of the above function and
calculate the entropy function defined as~\cite{0506177,0508042}:
\be \label{EFdef}
 {F(\vec u, \vec v, \vec e, \vec q, \vec p) \equiv 2\, 
\pi ( e_i \, q_i - f(\vec u, \vec v, \vec e, \vec p)) } \, ,
\ee
where,
\be
{ q_i\equiv {\p f \over \p e_i} }.
\ee
For the function $f$ given in eqn. (\ref{fbi}), we get:
\be
q_{1,3} = e_{1,3}\, \frac{v_2\,u_S\,u_R^2}
{2\sqrt{b\,v_1^2 - e_{1,3}^2\,u_S^2}}.
\ee
Using the above, the entropy function computed from 
eqn. (\ref{EFdef}) takes the form:
\bea \label{EF}
&&F(u_S, u_R,\wt u_R, v_1,v_2, q_1, q_3,p_2, p_4) 
= \frac{\pi}{2}\,(v_2-v_1)\,u_S  \xx
&& \, +\,
\pi\sqrt{b} \frac{v_1}{u_S}
\left[ \sqrt{4\,q^2_{1} 
+ bv_2^2\,u_R^4} - \sqrt{b}\,v_2 \, u_R^2 + \sqrt{4\,q^2_{3} 
+ bv_2^2\,u_R^{-4}} - \sqrt{b}\,v_2 \, u_R^{-2} \right] \xx
&& \,+ \,\sqrt{b} \frac{v_1}{4\,u_S}
\left[ \sqrt{u_S^2\,p^2_{2} 
+ 16\pi^2b\,v_2^2} - \pi \sqrt{b}\,v_2 \, \wt u_R^2 + \sqrt{u_S^2\,p^2_{4} 
+ 16\pi^2\,b\,v_2^2} - \pi \sqrt{b}\,v_2 \, \wt u_R^{-2} \right] \, .
\eea

Now, the attractor equations can be derived from:
\be \label{Ateq}
{\p F \over \p u_s}=0, \qquad {\p
F \over \p v_1}=0\,, \qquad  {\p
F \over \p v_2}=0\, , \qquad {\p
F \over \p u_R}=0\,, \qquad  {\p
F \over \p \wt u_R} =0\, .
\ee

In a generic case, the attractor equations may or may not be solvable.
In fact, in the case of four charge black hole we started with, 
it is difficult to solve the attractor equations analytically.
In particular, the attractor equations determining
$u_S,v_1,v_2$ are complicated. However, the attractor equations for
$u_R$ and $\wt u_R$ can still be written in terms of other moduli fields,
in a suggestive manner. For convenience, we give the 
result for $u_R$ in terms of the radial electric fields $e_1$ and $e_3$:
\be \label{Rsol}
{u_R}^4 = \frac{\left[\sqrt{b\,v_1^2 - e_3^2\,u_S^2} 
-\sqrt{b}\,v_1 \right]}{\left[\sqrt{b\,v_1^2 - e_1^2\,u_S^2} 
- \sqrt{b}\,v_1 \right]} \, ,
\ee
and for $\wt u_R$, we have:
\be \label{Rtsol}
{\wt u_R}^4 = \frac{\left[\sqrt{16\pi^2\,b\,v_2^2 + p_4^2\,u_S^2} 
-4\pi\,\sqrt{b}\,v_2 \right]}{\left[\sqrt{16\pi^2\,b\,v_2^2 + p_2^2\,u_S^2} 
-4\pi\,\sqrt{b}\,v_2 \right]} \, .
\ee
One can check that in the $b\rightarrow \infty$ limit, the standard 
result for $u_R$ and $\wt u_R$ is recovered~\cite{0508042} and these
results will be quoted below. Notice that the expressions for the entropy
function and final result for the moduli fields are explicitly duality
invariant.

\section{\bf Two Charge Black Holes}

As already mentioned,
it is in general hard to solve the attractor equations in the 
case when all four charges are present. Hence, we first set the 
magnetic charges to zero. We shall comment on the dyonic case in the
discussion section.

We are dealing with a small black hole, for which the 
area of the horizon and also the entropy vanishes to leading 
order~\cite{9504147}. Thus, in this case, the higher curvature 
corrections on the gravity side and even the Born-Infeld corrections on the
Maxwell side are supposed to be
important. It is known that upon the 
consideration of higher derivative corrections, a non-zero horizon emerges.
In this case, the attractor equations of the moduli fields are modified
and the solutions receive corrections. In this section, we shall analyze
the corrections to solutions of various moduli fields and to the entropy
from the Born-Infeld terms.

To start with, we do not include any higher curvature corrections on the
gravity side. In other words, we check whether the 
Born-Infeld terms independently play any role
near the horizon of a black hole. 
Therefore, we work with the Born-Infeld corrected effective
action given in eqn. (\ref{bi}). For convenience, we 
deal the case where  $q_1 = q_3 = q$, keeping
in mind that there are still two electric fields, but, with equal magnitudes
of charges and $p_2=p_4 =0$. That is, we only keep the charges 
associated with the momentum and winding numbers along the direction 
$x^9$ non-zero. 

To this end, terms in the Entropy function in eqn. (\ref{EF})
that are relevant for the electrically charged case are: 
\be \label{EFe}
F(u_S,v_1,v_2,q) = \pi\left[ \frac{u_S}{2}\,(v_2-v_1) 
+ 2\sqrt{b}\,\frac{v_1}{u_S}\left(\sqrt{4q^2 + bv_2^2} - \sqrt{b}\,v_2 \right)
\right]
\ee
Notice that we have already solved the attractor equation for the 
field $R$, which denotes the radius of the circle $S^1$, with the
result $u_R =1$. This solution is better found
before taking the Legendre transform, as it turns out to be a simpler
equation when written in terms of electric fields, as in eqn. (\ref{Rsol}). 
Using the relations in eqn.(\ref{Ateq}), one can derive the attractor
equations for various moduli fields from eqn. (\ref{EFe}). Solving
the attractor equations, one finds that (let us use $|q|$ to avoid negative
signs in the square root):
\be \label{r1}
v_2 = 2\,v_1 = \frac{4\,|q|}{\sqrt{3\,b}} \,  , \qquad
u_S^2 =  - \left(\frac{64\,b}{3}\right)^{1\over 2}\, |q| \, .
\ee
It turns out that the solution for the dilaton squared with a positive sign
is incompatible with rest of the attractor equations and the only solution
is when the sign is negative (neglecting the trivial $q=0$ solution). 
%Another way to put this inconsistency is that
%there is no solution to the attractor equations, for which we can consistently
%choose the same sign for all the charges . 
Thus, the solutions of 
attractor equations are not consistent, which means that the Born-Infeld terms
on their own are not enough to give a finite horizon and entropy to the 
black holes in this theory. It also signifies that black holes 
in this theory with only Born-Infeld terms do not have a nice extremal limit. 
This matches with the proposals in~\cite{0101083} about the non-existence
of extremal limit for electrically charged black holes in Born-Infeld 
theories. The case when magnetic charges are also included remains to
be checked.

In any case, let us appreciate the fact that 
solutions for the moduli fields $v_1$ and $v_2$,
though inconsistent, turn out to be of the order of 
$\alpha'$. Now, since there are $\alpha'$ corrections coming 
from the gravity side, one might wonder whether adding these terms 
has any effect on the result (\ref{r1}). Especially, since we have
set the magnetic charges to zero, comparing with the case in~\cite{0508042},
it is probably fine to assume that the string loop corrections (inverse 
square root of $u_S$) are insignificant in this limit 
and $\alpha'$ corrections are
important. 

Thus far, we have considered small black holes which have a vanishing horizon
and entropy if only the leading order terms in the action are present. The
addition of higher curvature terms changes this story. Thus, it is interesting
to consider higher curvature terms which can give same order contribution
to entropy as the Born-Infeld terms. With this in mind, we now 
include Gauss-Bonnet terms in the  
original action in eqn. (\ref{bi}) and look for solutions of 
attractor equations. Apriori, this may look strange since the
Gauss-Bonnet term is an $\alpha'$ correction, whereas the Born-Infeld
term considered is exact and contains terms 
to all orders in $\alpha'$. While, keeping only 
$\alpha'$ corrections on both sides is reasonable, solving
the attractor equations order by order in $\alpha'$ is tedious. 
Thereupon, we resort to an exact treatment on the Born-Infeld side. 
Besides,
the final result for entropy with Born-Infeld terms turns out to be of the 
order of $\alpha'$, as seen from eqn. (\ref{r1}).

In~\cite{0508042}, the corrections to the entropy from the Gauss-Bonnet terms  
have already been computed and we use those results here. The terms
added to the effective action of heterotic string theory were of the 
form : 
\be \label{ec1}
\Delta\LL =  C\,{{S}\over 16\pi}\, 
\left\{ R_{G\mu\nu\rho\sigma} R_G^{\mu\nu\rho\sigma} 
- 4 R_{G\mu\nu} R_G^{\mu\nu}
+ R_G^2
\right\} \, ,
\ee
where $R_{G\mu\nu\rho\sigma}$ denotes the Riemann tensor computed using
the string metric $G_{\mu\nu}$. The Gauss-Bonnet parameter is given as,
$C= \alpha'/16$ and  is equal to $1$ according to the present notations. 
Due to the above term, solutions for various moduli fields
get modified as (setting $p_2 = p_4 =0$)~\cite{0508042}: 
\be  \label{senGB}
v_1 = v_2  = 8\, , \qquad 
u_S =  \sqrt{q_1q_3} \, .
\ee
In fact, the result for entropy gets corrected as: 
\be \label{SqC}
S_{BH}  = 8\pi \, \sqrt{C\,q_1q_3} \, ,
\ee 
with the values of other moduli remaining unchanged.

In the present case, including the Gauss-Bonnet terms, the
correction to the entropy function can be calculated 
with the following changes in the functions $f$ and $F$:
\be \label{ec2}
 \Delta f = -2 \, u_S \,, \quad 
 \quad {\Delta F = 4\, \pi \, u_S} \, ,
\ee
and the relation between $q_i$ and $e_i$ unchanged. Solving the 
attractor equations for $F + \Delta F$ following from eqn.(\ref{EFe}) and
(\ref{ec2}), we have:
\bea \label{r3}
&& v_1 = C\,\frac{8}{3}\left(2\, 
+ \,\sqrt{1 - \frac{3\,q^2}{16\,b\,C^2}}\right)\, ,
\qquad v_2 = C\,\frac{8}{3}\left(1 + 2\,\sqrt{1 - 
\frac{3\,q^2}{16\,b\,C^2}}\right)\,,\xx
&& u_S = C\,\sqrt{\frac{32\,b}{3}}
\left(1 \,- \,\sqrt{1 - \frac{3\,q^2}{16\,b\,C^2}} 
\right)^{1\over 2}. 
\eea
There are several things to note from the above solution.
First of all, it is extremely interesting to note that now there are 
consistent solutions of the attractor equations, 
where the radii of $AdS_2$, $S^2$ and the dilaton, get 
fixed in terms of the charge $q$. This signifies the presence of attractor
mechanism in this theory. Tracing back, one can see what 
happens to eqns. (\ref{r3}), if the Gauss-Bonnet term is set to zero.  
This leads us to a very important conclusion that the Gauss-Bonnet
contribution provides a bound for the existence of extremal
configurations and for the presence of attractor mechanism 
in this theory. This in turn brings about the importance
of $\alpha'$ corrections from the near horizon analysis. 
In fact, if the above analysis could be repeated in higher 
dimensions, with the higher Gauss-Bonnet densities, then, there might be a 
suggested bound on the charges, from the coefficients of 
Lovelock terms to each order~\cite{0511306}, putting a bound 
on the existence of extremal configurations.

Now as expected, the entire correction to the radius of $AdS_2$ and
$S^2$ in eqn. (\ref{r3}) is of the order of $\alpha'$. This is because 
the combination $b\,C^2 = \frac{1}{4^5\,\pi^2}$, is dimensionless. 
Thus, to this order, 
various moduli fields receive corrections not just from the Gauss-Bonnet 
part~\cite{0508042}, but also from the Born-Infeld part. Since these
corrections to the horizon are of the same order, per se, the Born-Infeld
corrections cannot be ignored. The aforementioned bound on the existence
of extremal configurations in this theory is:
\be
q \: < \: 8\sqrt{3}\,\pi.
\ee
As long as this bound is satisfied, the solutions in eqn. (\ref{r3}) 
are consistent and attractor mechanism works. 

One can draw further conclusions from
the solutions for the moduli fields given in eqn. (\ref{r3}). One 
notices that the radii of $AdS_2$ and $S^2$ are not equal, unlike the
correction given by a Gauss-Bonnet terms~\cite{0508042}.
Such a correction to the radii from the Born-Infeld terms
means that the near horizon geometry will no longer be 
conformally flat.
This further signifies that the cosmological constant is also non-zero. 
Unbroken supersymmetry, if present would 
require that the radii of $AdS_2$ and $S^2$ should be equal. In fact, this
is what is expected if one uses a supersymmetric attractor mechanism.
Thus, the solutions in eqn. (\ref{r3}) should be non-supersymmetric, with
the sign of $q^2$ chosen to be positive. 
This is consistent with the present formalism, where neither the solution 
nor the method of deriving the entropy function requires supersymmetry to be present.
In the present example, taking $q^2$ to be negative is inconsistent and 
does not give any result in the corresponding non-BPS case. 

Let us also note the fact that, if we take $\alpha' \rightarrow 0$ limit 
in the solutions in eqn. (\ref{r3}), one clearly recovers the standard
result, that $AdS_2$ and $S^2$ have the same radii, which is zero and 
the naked singularity reappears. Furthermore, the dilaton diverges, as
is expected from the results in~\cite{0508042}(since we did not include
magnetic charges).

Let us now turn to the Born-Infeld corrections to the entropy of 
the extremal black holes in this theory.
Using the solutions of the moduli fields in the entropy function gives
us the entropy of the near horizon configuration in this theory. Without
dwelling in to calculational details, let us present the final result. The 
entropy comes out to be finite and gets corrected as:
\bea \label{SBHfinal}
S_{BH} &=& 4\pi\,\frac{u_S\,v_1}{4} \:
= \frac{32\pi}{3}\,\sqrt{\frac{2}{3}}\, \sqrt{bC^3}\,
\left( 1 \,- \sqrt{1\,- \,\frac{3\,q^2}{16\,b\,C^2}} \right)^{1\over 2}\,
\left( 2 \,+ \,\sqrt{1 - \frac{3\,q^2}{16\,b\,C^2} } \right)\, ,\\
&=& \frac{1}{12}\sqrt{\frac{2\alpha'}{3}}\,
\left( 1 \,- \sqrt{1\,- \,(8\sqrt{3}\pi)^2\,q^2} \right)^{1\over 2}\,
\left( 2 \,+ \,\sqrt{1 - (8\sqrt{3}\pi)^2\,q^2} \right).
\eea
For an extremal black hole in the theory governed by the action in 
eqn. (\ref{bi}), with Gauss-Bonnet terms in addition, the entropy comes out
to be as in eqn. (\ref{SBHfinal}). It is worth mentioning that the 
characteristic dependence of the Born-Infeld parameter on the entropy of 
extremal black holes has not been calculated before, either from the 
near horizon analysis or from the full solution. Although, the above results
were derived in a rather simple set up, the importance increases, 
considering the fact that a similar calculation with all four
charges turned on is already not possible analytically. 

Let us look at the Born-Infeld contribution to the entropy of
the extremal black holes in this theory. To ensure that our results are
correct, it is important to check that taking appropriate
limits in eqn. (\ref{SBHfinal}) and comparing it with the known results,
we get the right answers.
For instance, one can take the $b\rightarrow \infty$ limit which 
corresponds to recovering the results in the Maxwell regime considered
in~\cite{0508042}. In this limit, keeping the Gauss-Bonnet parameter
$C$ intact, for the entropy, one recovers the result 
$S_{BH} = 8\pi \, \sqrt{C\,q_1q_3}  $, same as in eqn. (\ref{SqC}).
Furthermore, 
in eqn. (\ref{SBHfinal}), taking the 
$\alpha' \rightarrow 0$ limit, one recovers the result for the entropy
of small black holes, which vanishes due to the vanishing of horizon
area. Moreover, in the results for the moduli fields in eqn. (\ref{r3}),
one can  check what happens if the charge $q$ is not present. 
In this case, one recovers the 
results of~\cite{0508042}, where the correction to the horizon due to
the Gauss-Bonnet terms is evident. Thus, our results for the moduli 
fields and the entropy have the correct limits and show the additional
corrections due to the Born-Infeld terms in the effective action. The
above analysis also shows the importance of adding higher curvature 
terms for the attractor mechanism to work in this theory.

\section{\bf Discussion}

In this letter, we studied the effect of Born-Infeld on the entropy 
of small black holes in heterotic
string theory in four dimensions and the attractor mechanism, 
using the entropy function formalism.
We found that the near horizon geometry
gets corrections from the Born-Infeld terms, which are of the
same order as the Gauss-Bonnet contribution. The resulting 
entropy comes out to be finite and shows additional corrections due to
the Born-Infeld terms. However, these corrections are inconsistent without
the addition of Gauss-Bonnet terms. Moreover, the attractor mechanism 
works consistently only after the addition of such higher curvature terms
to the effective action. It is thus important to include 
all possible corrections to the horizon geometry at a given order in
$\alpha'$. In this work, we restricted ourselves to
only two charge black holes in four dimensions.
The case of a dyonic black hole is interesting, although the calculations
involve solving coupled higher order algebraic equations. We have not been
able to find any analytic solutions in this case. It is also important to
include other types of terms in the calculation. For instance,   
it should be interesting to explore what happens when the 
S-duality invariant axionic coupling of the Born-Infeld and 
Gauss-Bonnet terms is considered in this theory.    

It is to be noted that, in the entropy function formalism, 
it is implicit that all
the scalar fields at the horizon take constant values. There exist
other approaches
to non-supersymmetric attractor mechanism, which
make explicit use of the general solution and equations of 
motion~\cite{0507096}. In this picture, extremization of an 
effective potential, 
gives rise to the necessary attractor equations.
In addition, there is sufficiency condition, which states that the
extremum should also be a minimum. Furthermore,
using a perturbative approach to study the corrections
to the scalar fields and taking the backreaction corrections in to the
metric, it is possible to show that the scalar fields are indeed drawn
to their fixed values at the horizon. Here, the requirements are the
existence of a {\it double degenerate horizon solution}, as in a 
Reissner-Nordstrom black hole. 

Recently, the connection 
between the 'entropy function' and the 
effective potential approach has been detailed in~\cite{0601016}
and non-supersymmetric attractors in $R^2$ gravities have been
discussed in~\cite{Chandrasekhar:2006kx}. 
In the absence of supersymmetry, it cannot be guaranteed that the
extremum of the effective potential will also be a minimum and hence,
this must be checked on a case to case basis.

The analysis in this paper provides weightage to the 
existence of a consistent extremal limit for electrically charged 
black holes in Born-Infeld-Gauss-Bonnet theories. Motivated by the near 
horizon analysis in this work, it should be possible to show that there
is no attractor mechanism for black holes in purely 
Einstein-Born-Infeld theories coupled to moduli fields. One might hope to
see this feature in the non-existence of an extremum of the
the effective potential of the theory. It is important to check this
aspect for confirmation, using other methods of studying attractor 
mechanism. Finally, attractor
mechanism seems to work only after the addition of other higher curvature
terms from the gravity side, for example the Gauss-Bonnet term in this
case. It is thus an extremely interesting and important problem 
to check these issues using the equations of motion explicitly
for Einstein-Born-Infeld and Gauss-Bonnet systems coupled to 
moduli fields in four and higher dimensions, either by analytic or numerical
methods~\cite{0507096,Chandrasekhar:2006kx}.

It is also important to remember that the formalism prescribed 
in~\cite{0508042} does not use any supersymmetry to derive attractor
equations for the moduli fields and to calculate the entropy. Further,
the Gauss-Bonnet term added to the action did not come with its supersymmetric
partner. Similarly, in this work, we have only considered the bosonic
part of the Born-Infeld terms to the action. We neither added any higher 
derivative terms on the Maxwell side or considered a supersymmetric 
Born-Infeld lagrangian. In fact, it is not clear whether it is 
possible to get exact analytical results when other types of 
gauge-gravitational terms are included in
the action. The fact that an exact calculation to all orders
in $\alpha'$ for the Born-Infeld part is possible for two charge black 
holes, owes to the simplicity of the `entropy function' formalism. It 
should be interesting to include other possible terms appearing in Born-Infeld
theories and check their dependence on entropy. All these issues remain to be
pursued and we hope to return to them in future.

\vskip 1.0cm

\begin{center}
{\bf Acknowledgements}
\end{center}

I wish to thank M. Alishahiha and M. M. Sheikh-Jabbari for helpful
discussions and suggestions, and S. Mukherji for a useful communication. 
I am grateful to A. Sen for comments on the preliminary draft.

\end{document}